\documentclass[10pt,twocolumn,letterpaper]{article}

\usepackage{iccv}
\usepackage{times}
\usepackage{epsfig}
\usepackage{graphicx}
\usepackage{amsmath}
\usepackage{amssymb}

\usepackage{adjustbox}
\usepackage{gensymb}

\usepackage[breaklinks=true,bookmarks=false]{hyperref}
\hypersetup{
    colorlinks=true,
    linkcolor=blue,
    filecolor=magenta,      
    urlcolor=red,
}
\iccvfinalcopy % *** Uncomment this line for the final submission

\def\iccvPaperID{****} % *** Enter the ICCV Paper ID here

\ificcvfinal\fi

\begin{document}

%%%%%%%%% TITLE
\title{ MIA-COV19D: COVID-19 Detection through 3-D Chest CT Image Analysis}

\author{

Dimitrios Kollias\\
University of Greenwich, UK\\
{\tt\small D.Kollias@greenwich.ac.uk} \\

\and 

Anastasios Arsenos \\  
National Technical University\\ Athens, Greece \\

\and
Levon Soukissian \\
GRNET National Infrastructures\\ Research \&  Technology, Greece \\
\and
Stefanos Kollias \\ 
National Technical University\\ Athens, Greece     \\

}

\maketitle
\ificcvfinal\thispagestyle{empty}\fi

%%%%%%%%% ABSTRACT
\begin{abstract}
   
   Early and reliable COVID-19  diagnosis based on chest 3-D  CT scans can assist medical specialists in vital circumstances.
   Deep learning methodologies constitute a main approach  for chest CT scan analysis and disease prediction. However, large annotated databases are necessary for developing deep learning models that are able to provide COVID-19 diagnosis  across various medical environments in different countries. 
   Due to privacy issues, publicly available COVID-19 CT datasets are highly difficult to obtain, which hinders the research and development of AI-enabled diagnosis methods of COVID-19 based on CT scans. 
   
   In this paper we present the COV19-CT-DB database which is annotated for COVID-19, 
   consisting of about 5,000 3-D CT scans, We have split the database in training, validation and test datasets. The former two datasets can be used for  training and validation of machine learning models, while the latter will be used for evaluation of the developed models. We also present a deep learning approach, based on a CNN-RNN network and report its performance on the COVID19-CT-DB database.

\end{abstract}

%%%%%%%%% BODY TEXT
\section{Introduction}

The Coronavirus Disease 2019 SARS-CoV-2 (COVID-19) has become a global pandemic with an exponential growth and mortality rate. 
The virus is harbored most commonly with little or no symptoms, but can also lead to a rapidly progressive and often fatal pneumonia \cite{sohrabi2020world,lai2020asymptomatic,hoehl2020evidence}. 

It has become important to detect affected people as early as possible and isolate them to stop further spreading of the virus.
Various methods have been proposed to diagnose COVID-19, containing a variety of medical imaging techniques, blood tests and PCR. 

COVID-19 pandemic has a very severe impact on the respiratory as well as other systems of the human body. Thus, medical imaging features of chest radiography is found to be useful for rapid COVID-19 detection. The imaging features of the chest can be obtained through medical imaging modalities like CT (Computed Tomography) scans. CT images can be used for  precise COVID-19 detection \cite{alizadehsani2021risk}.

They provide: a) 3-D view formation of organs; CT scans provide a more detailed overview of the internal structure of lung parenchyma due to lack of overlapping tissues, b) convenient examination of disease and its location; CTs  provide a window into pathophysiology that could shed light on several stages of disease detection and evolution. Radiologists report COVID-19 patterns of infection  with typical features including ground glass opacities in the lung periphery, rounded opacities, enlarged intra-infiltrate vessels, and later more consolidations that are a sign of progressing critical illness.

At the time of CT scan recording, several slices are captured from each person suspected of COVID-19. The large volume of CT scan images calls for a high workload on physicians and radiologists to diagnose COVID-19. Taking this into account and also the rapid increase in number of new and suspected COVID-19 cases, it is evident that there is a need for using machine and deep learning for detecting COVID-19 in CT scans. 

Such approaches require data to be trained on. Therefore, a few databases have been developed consisting of CT scans. However, new data sets with large numbers of 3-D CT scans are needed, so that researchers can train and develop COVID-19 diagnosis systems and trustfully evaluate their performance.    

The current paper presents a baseline approach for the Competition part of the Workshop “AI-enabled Medical Image Analysis Workshop and Covid-19 Diagnosis Competition (MIA-COV19D)” which occurs in conjunction with the International Conference on Computer Vision (ICCV) 2021 in Montreal, Canada, October 11- 17, 2021. 

The MIA-COV19D AI-enabled Medical Image Analysis (MIA) Workshop emphasizes on radiological quantitative image analysis for diagnosis of diseases. The focus is placed on Artificial Intelligence (AI), Machine and Deep Learning (ML, DL) approaches that target effective and adaptive diagnosis, as well as on  approaches that enforce trustworthiness and create justifications of the decision making process.

The COV19D Competition is based on a new large database of chest CT scan series that is manually annotated for Covid-19/non-Covid-19 diagnosis. The training and validation partitions along with their annotations are provided to the participating teams to develop AI/ML/DL models for Covid-19/non-Covid-19 prediction. Performance of approaches will be evaluated on the test set.

The COV19-CT-DB is a new large database with about 5,000 3-D CT scans, annotated for COVID-19 infection. 

The rest of the paper is as follows. Section 2 presents former work on which the presented baseline has been based. Section 3 presents the database created and used in the Competition. The ML approach and the pre-processing steps  are described in Section 4. The obtained results, are presented in Section 5. Conclusions and future work are described in Section 6.

\section{Related Work}

In \cite{khadidos2020analysis} a CNN plus RNN network was used, taking as input CT scan images and discriminating
between COVID-19 and non-COVID-19 cases. 

In \cite{li2020coronavirus}, the authors employed a variety of 3-D ResNet models for detecting COVID-19 and distinguishing it from other common pneumonia (CP) and normal cases, using  volumetric 3-D CT scans. 

In \cite{wang2020weakly}, a weakly supervised deep learning framework was suggested using 3-D CT volumes for COVID-19 classification and lesion localization. A pre-trained UNet was utilized for segmenting the lung region of each CT scan slice; the latter was fed into a 3-D DNN that provided the classification outputs.  

The presented approach is based on a CNN-RNN architecture that performs 3-D CT scan analysis. The method follows our previous work \cite{Tailor, springer, cis, ijait} on developing deep neural architectures for  predicting COVID-19, as well as neurodegenerative and other \cite{iet, cis, 48aposcholarmou, mdpi} diseases and medical situations. 

These architectures have been applied for: a) prediction of Parkinson’s, based on datasets of MRI and DaTScans, either created in collaboration with the Georgios Gennimatas Hospital (GGH) in Athens \cite{cis}, or provided by the PPMI study sponsored by M. J. Fox for Parkinson’s Research \cite{iet}, b) prediction of COVID-19, based on CT chest scans, scan series, or x-rays, either collected from the public domain, or aggregated in collaboration with the Hellenic Ministry of Health and The Greek National Infrastructures for Research and Technology \cite{Tailor}.

\section{The COV19-CT-DB Database}

\begin{figure*}[h!]
\centering
\adjincludegraphics[height=2.7cm]{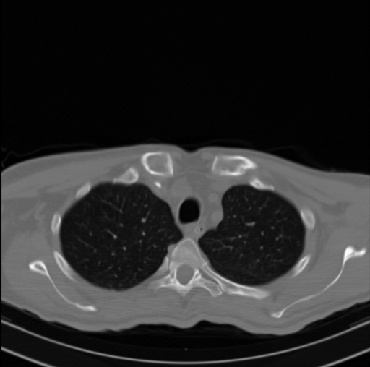}
\adjincludegraphics[height=2.7cm]{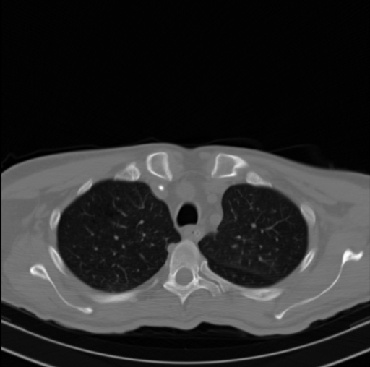}
\adjincludegraphics[height=2.7cm]{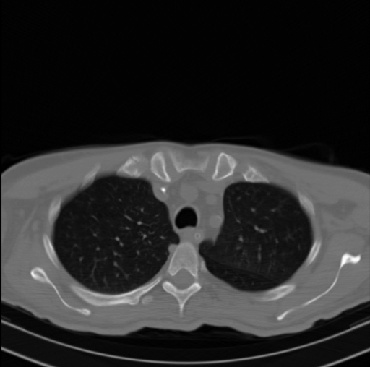}
\adjincludegraphics[height=2.7cm]{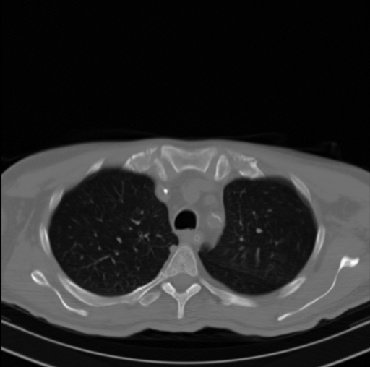}
\adjincludegraphics[height=2.7cm]{non_covid_cluster_17/cam_series_1_slice_16.jpg}
\adjincludegraphics[height=2.7cm]{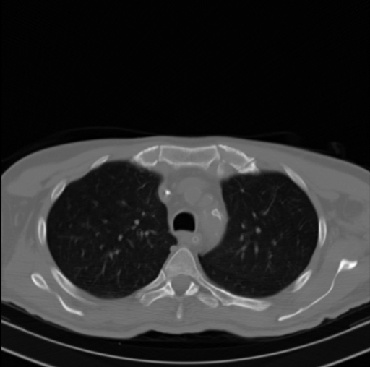}
\adjincludegraphics[height=2.7cm]{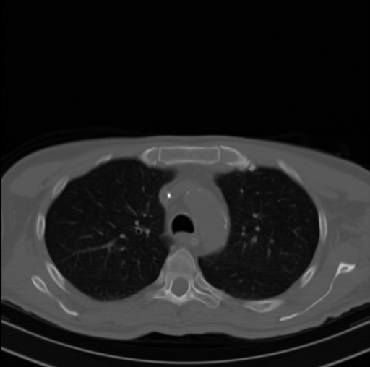}
\adjincludegraphics[height=2.7cm]{non_covid_cluster_17/cam_series_1_slice_18.jpg}
\adjincludegraphics[height=2.7cm]{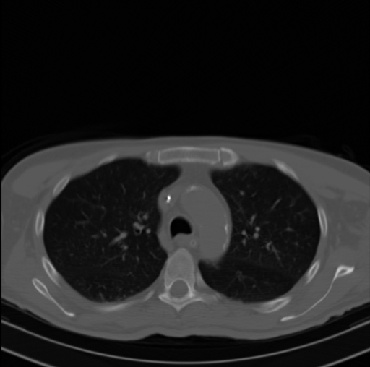}
\adjincludegraphics[height=2.7cm]{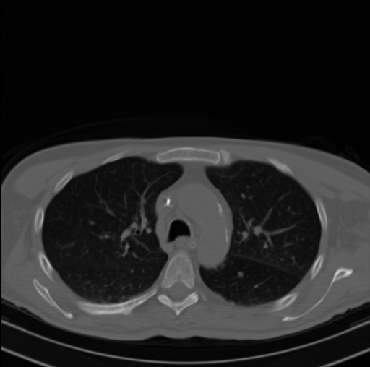}
\adjincludegraphics[height=2.7cm]{non_covid_cluster_17/cam_series_1_slice_20.jpg}
\adjincludegraphics[height=2.7cm]{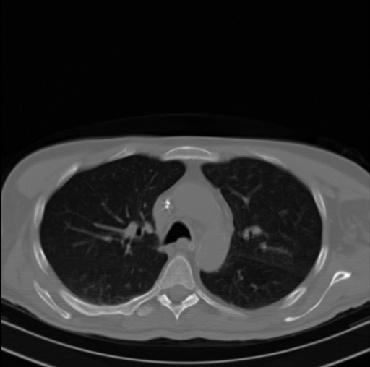}
\adjincludegraphics[height=2.7cm]{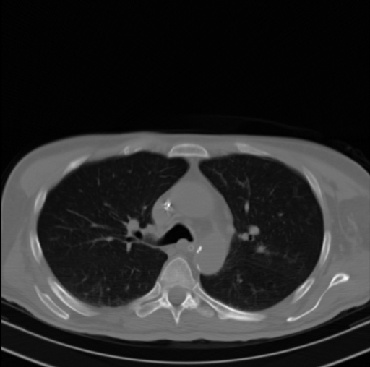}
\adjincludegraphics[height=2.7cm]{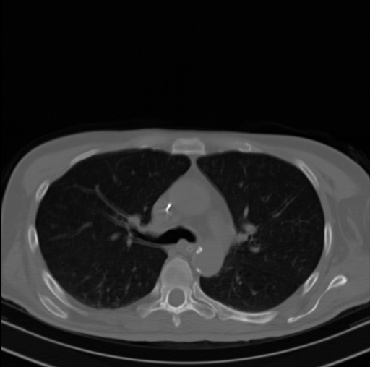}
\adjincludegraphics[height=2.7cm]{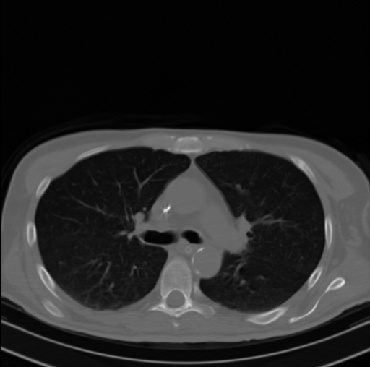}
\adjincludegraphics[height=2.7cm]{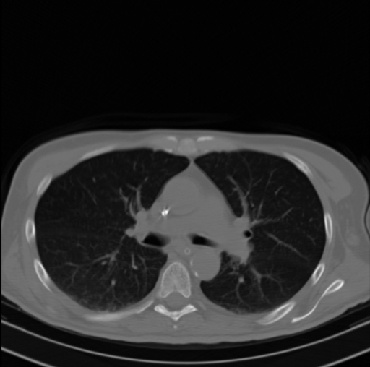}
\adjincludegraphics[height=2.7cm]{non_covid_cluster_17/cam_series_1_slice_25.jpg}
\adjincludegraphics[height=2.7cm]{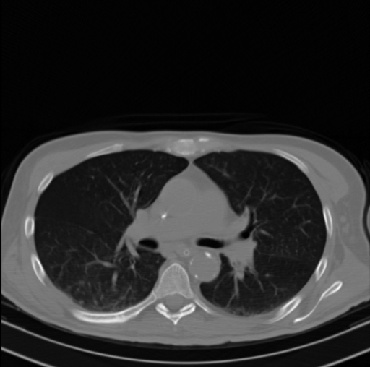}
\adjincludegraphics[height=2.7cm]{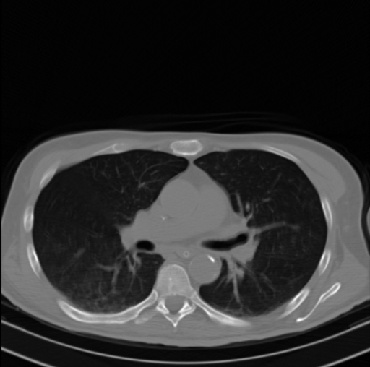}
\adjincludegraphics[height=2.7cm]{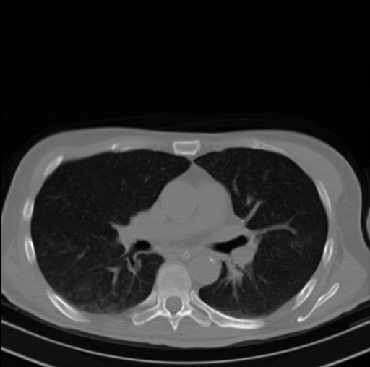}
\adjincludegraphics[height=2.7cm]{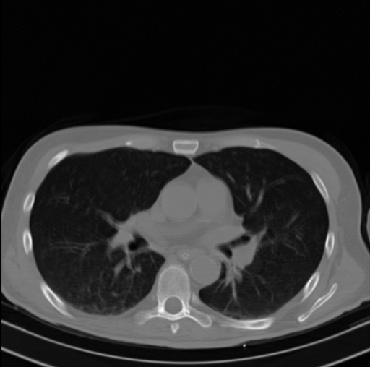}
\adjincludegraphics[height=2.7cm]{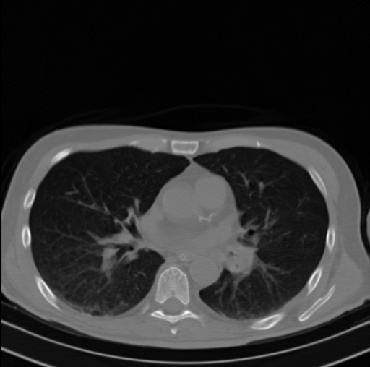}
\adjincludegraphics[height=2.7cm]{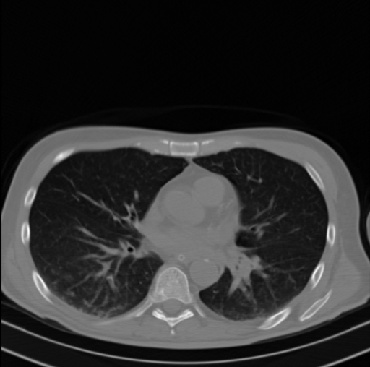}
\adjincludegraphics[height=2.7cm]{non_covid_cluster_17/cam_series_1_slice_31.jpg}
\adjincludegraphics[height=2.7cm]{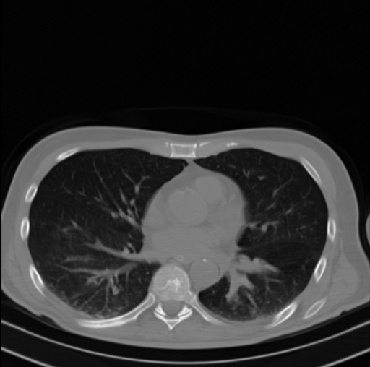}
\adjincludegraphics[height=2.7cm]{non_covid_cluster_17/cam_series_1_slice_32.jpg}
\adjincludegraphics[height=2.7cm]{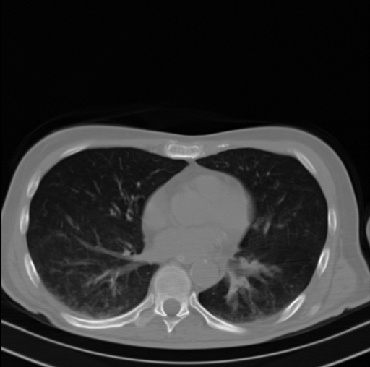}
\adjincludegraphics[height=2.7cm]{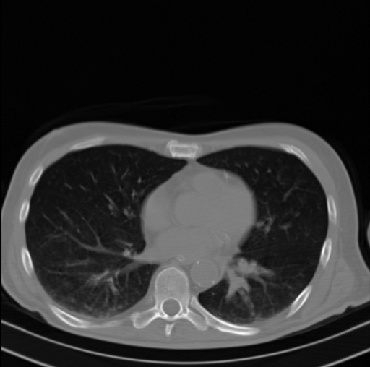}
\adjincludegraphics[height=2.7cm]{non_covid_cluster_17/cam_series_1_slice_34.jpg}
\adjincludegraphics[height=2.7cm]{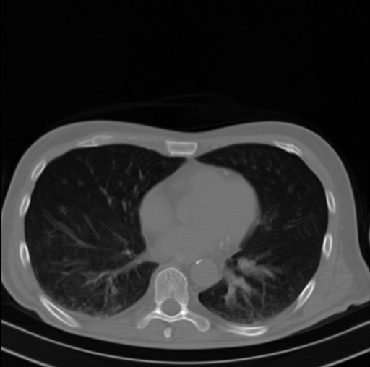}
\caption{Slices from a non COVID-19 CT scan.}
\label{full_non_covid_ct_scan}
\end{figure*}

\begin{figure*}[h!]
\centering
\adjincludegraphics[height=2.7cm]{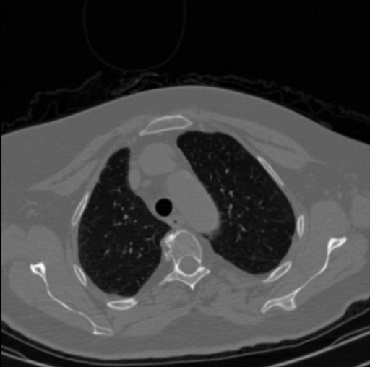}
\adjincludegraphics[height=2.7cm]{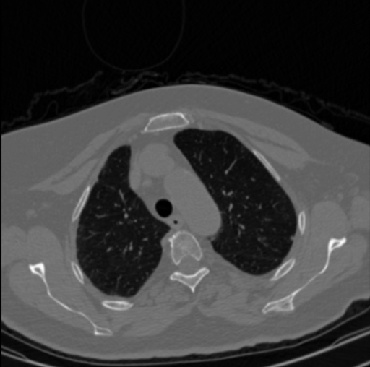}
\adjincludegraphics[height=2.7cm]{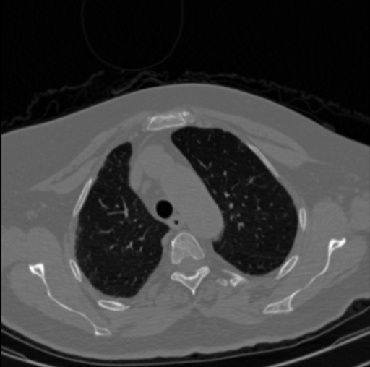}
\adjincludegraphics[height=2.7cm]{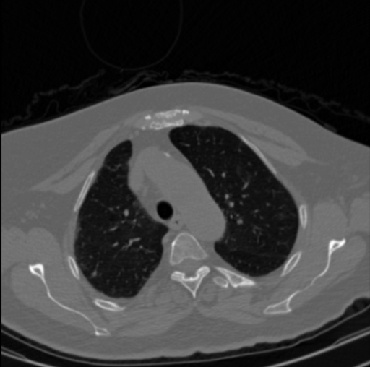}
\adjincludegraphics[height=2.7cm]{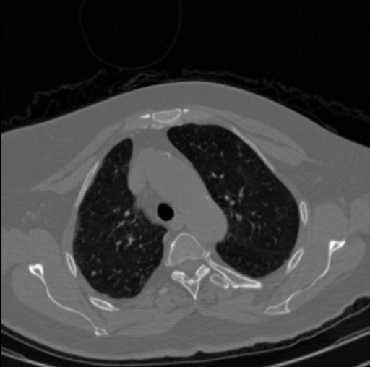}
\adjincludegraphics[height=2.7cm]{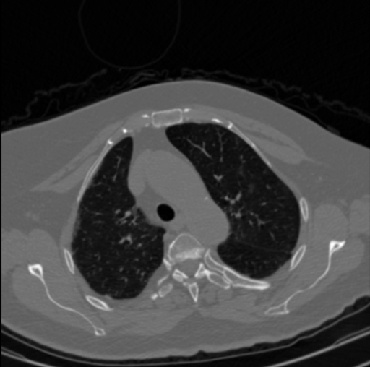}
\adjincludegraphics[height=2.7cm]{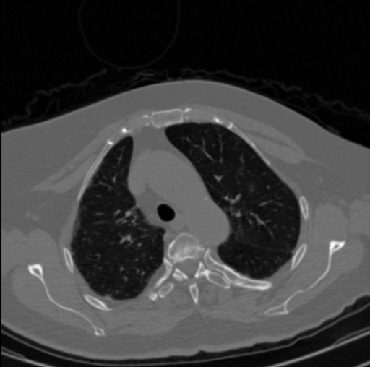}
\adjincludegraphics[height=2.7cm]{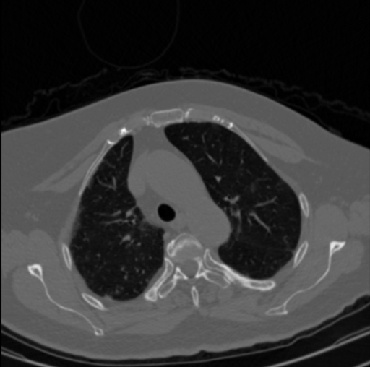}
\adjincludegraphics[height=2.7cm]{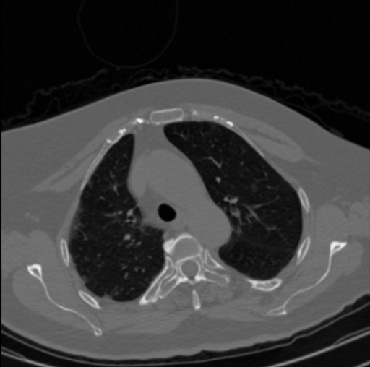}
\adjincludegraphics[height=2.7cm]{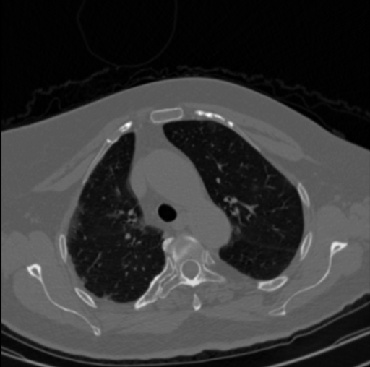}
\adjincludegraphics[height=2.7cm]{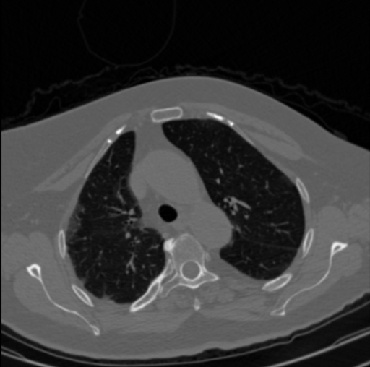}
\adjincludegraphics[height=2.7cm]{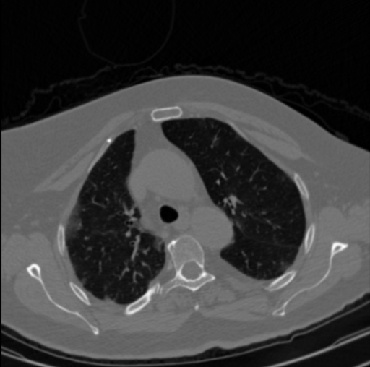}
\adjincludegraphics[height=2.7cm]{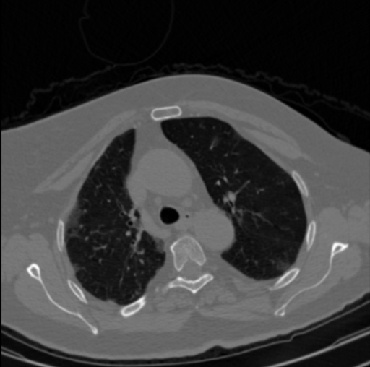}
\adjincludegraphics[height=2.7cm]{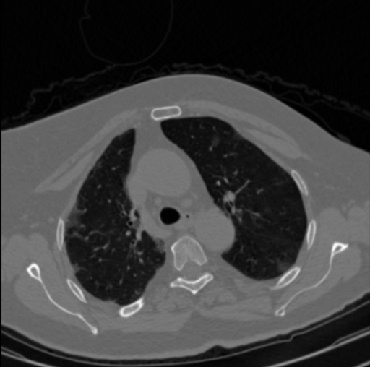}
\adjincludegraphics[height=2.7cm]{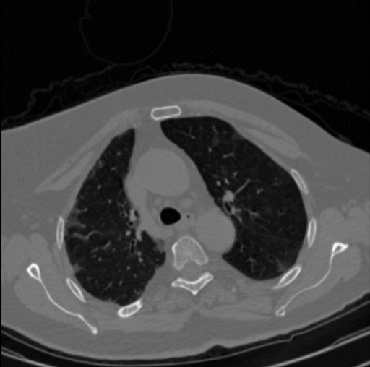}
\adjincludegraphics[height=2.7cm]{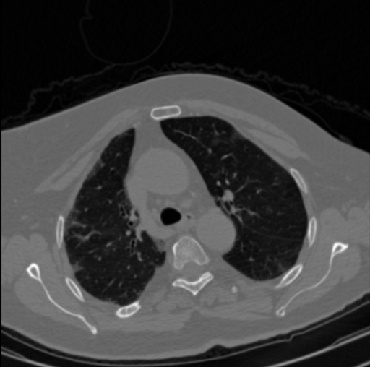}
\adjincludegraphics[height=2.7cm]{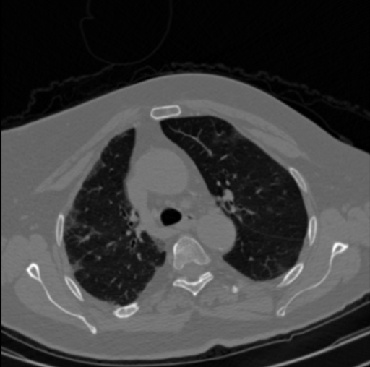}
\adjincludegraphics[height=2.7cm]{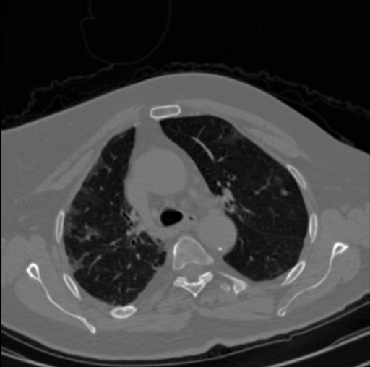}
\adjincludegraphics[height=2.7cm]{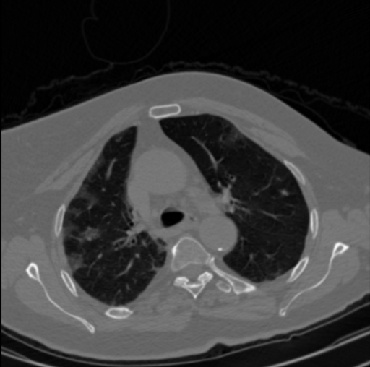}
\adjincludegraphics[height=2.7cm]{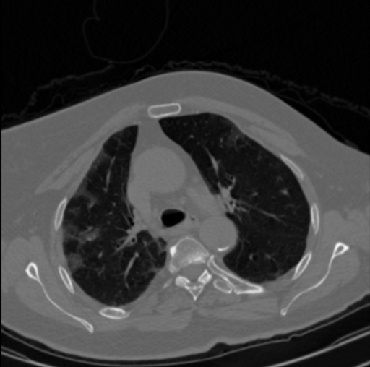}
\adjincludegraphics[height=2.7cm]{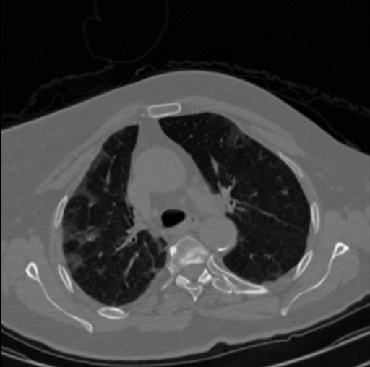}
\adjincludegraphics[height=2.7cm]{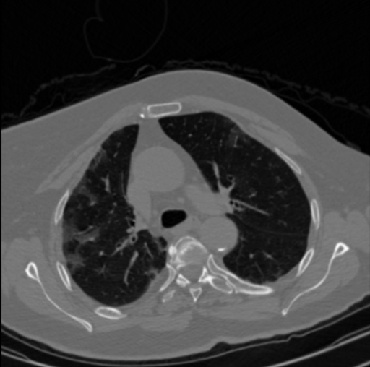}
\adjincludegraphics[height=2.7cm]{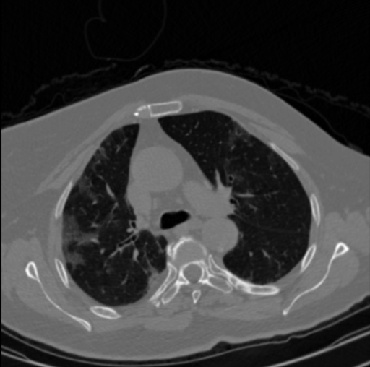}
\adjincludegraphics[height=2.7cm]{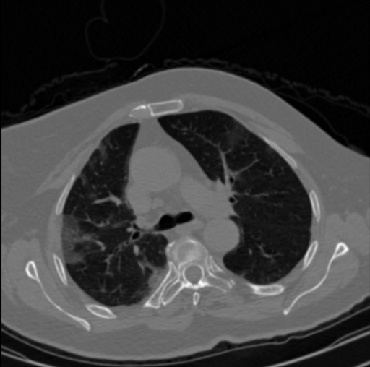}
\adjincludegraphics[height=2.7cm]{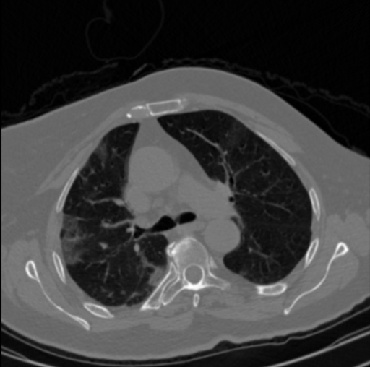}
\adjincludegraphics[height=2.7cm]{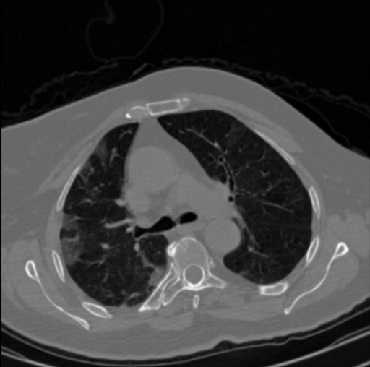}
\adjincludegraphics[height=2.7cm]{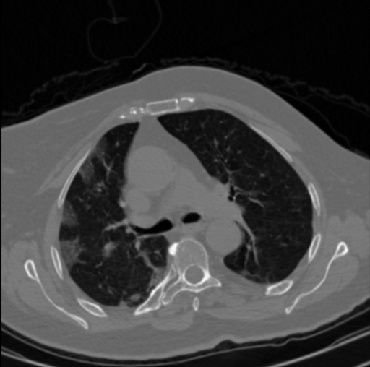}
\adjincludegraphics[height=2.7cm]{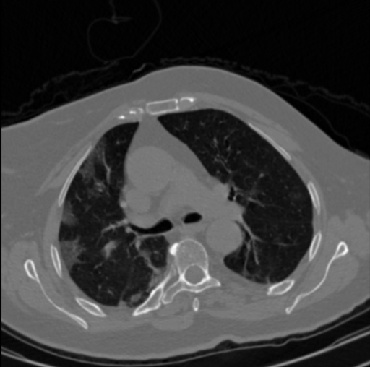}
\adjincludegraphics[height=2.7cm]{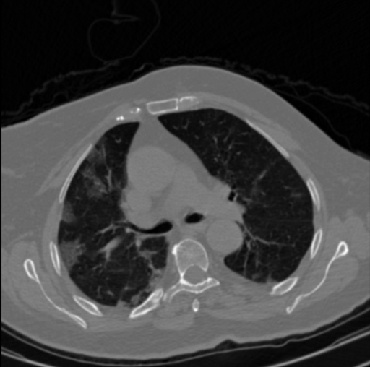}
\adjincludegraphics[height=2.7cm]{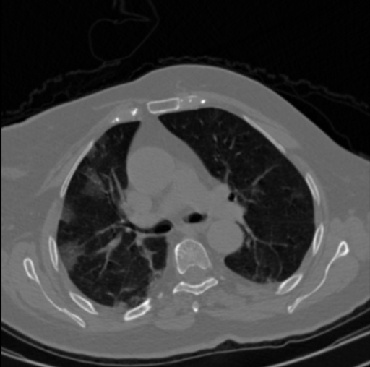}
\adjincludegraphics[height=2.7cm]{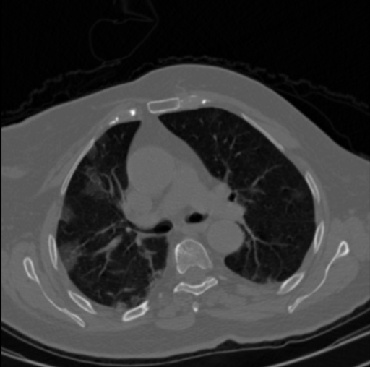}
\adjincludegraphics[height=2.7cm]{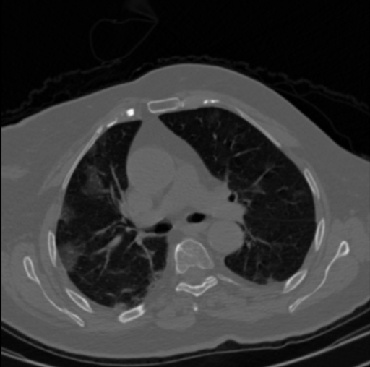}
\adjincludegraphics[height=2.7cm]{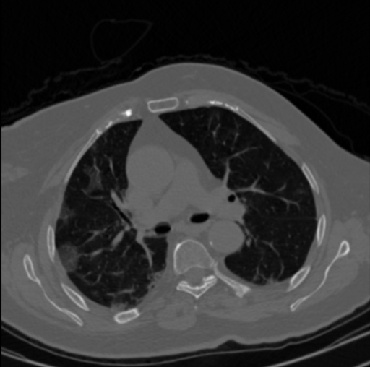}
\adjincludegraphics[height=2.7cm]{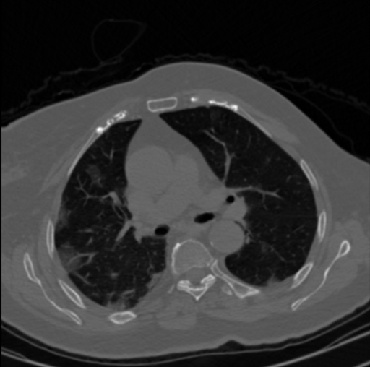}
\adjincludegraphics[height=2.7cm]{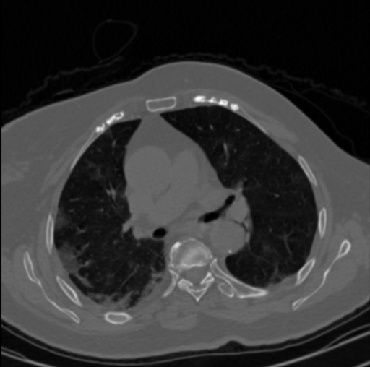}
\adjincludegraphics[height=2.7cm]{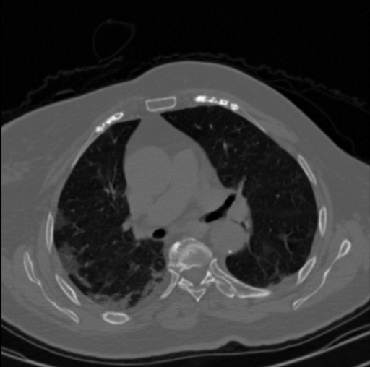}
\adjincludegraphics[height=2.7cm]{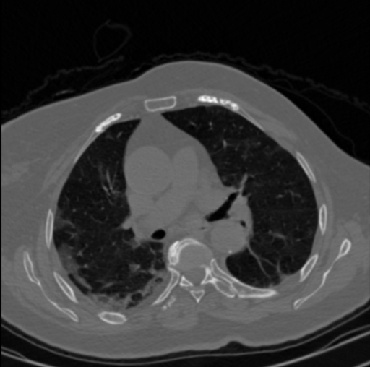}
\adjincludegraphics[height=2.7cm]{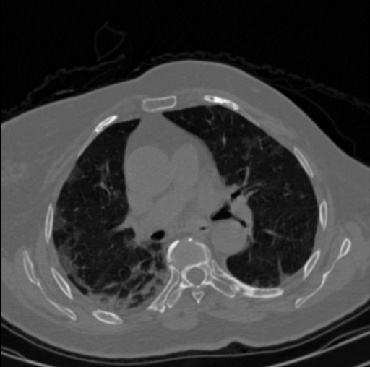}
\adjincludegraphics[height=2.7cm]{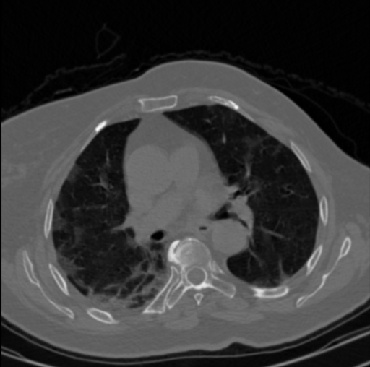}
\adjincludegraphics[height=2.7cm]{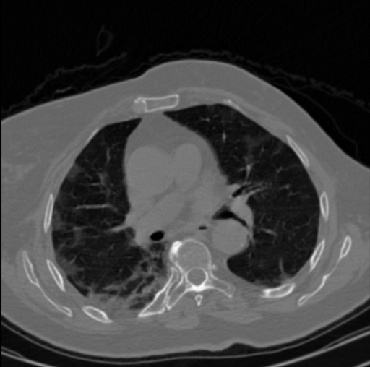}
\adjincludegraphics[height=2.7cm]{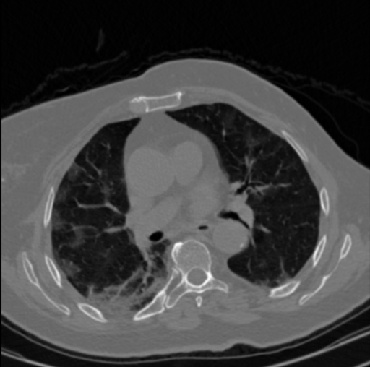}
\adjincludegraphics[height=2.7cm]{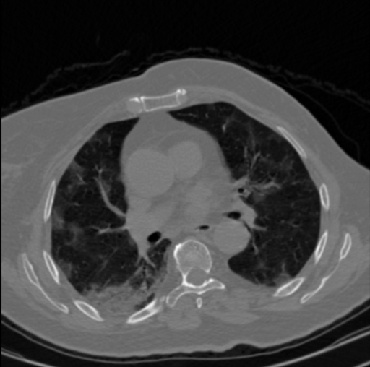}
\caption{Slices from a COVID-19 CT scan.}
\label{full_covid_ct_scan}
\end{figure*}

The COVID19-CT-Database (COV19-CT-DB) consists of chest CT scans that are annotated for the existence of COVID-19. Data collection was conducted in the period from September 1, 2020 to  March  31,  2021. Data were aggregated from many hospitals, containing  anonymized  human  lung CT scans with signs of COVID-19 and  without  signs  of  COVID-19. Figure \ref{full_non_covid_ct_scan} shows some CT slices from a non-COVID-19 case and Figure \ref{full_covid_ct_scan} some CT slices from a COVID-19 case.

The COV19-CT-DB database consist of about 5000  chest CT scan series, which correspond to a high number of patients ($>$1000) and subjects ($>$2000).  Annotation of each CT slice has been performed by 4 very experienced (each with over 20 years of experience) medical experts; two radiologists and two pulmonologists. Labels provided by the 4 experts showed a high degree of agreement (around 98\%). 

One difference of COV19-CT-DB from other existing datasets is its annotation by medical experts (labels have not been created as a result of just positive RT-PCR testing).

Each of the 3-D scans includes different number of slices, ranging from 50 to 700. 
The database has been split in training, validation and testing sets. 

The training set contains, in total, 1560 3-D CT scans. These include 690 COVID-19 cases and  870 Non-COVID-19 cases. The validation set consists of 374 3-D CT scans.  165 of them represent COVID-19 cases and  209 of them represent Non-COVID-19 cases. Both include different numbers of CT slices per CT scan, ranging from 50 to 700.

\section{The Deep Learning Approach}

\subsection{3-D Analysis and COVID-19 Diagnosis}

The input sequence is a 3-D signal, consisting of a series of chest CT slices, i.e., 2-D images, the number of which is varying, depending on the context of CT scanning. The context is defined in terms of various requirements, such as the accuracy asked by the doctor who ordered the scan, the characteristics of the CT scanner that is used, or the specific subject’s features, e.g., weight and age.

\begin{figure*}[h!]
\centering
\adjincludegraphics[height=11cm]{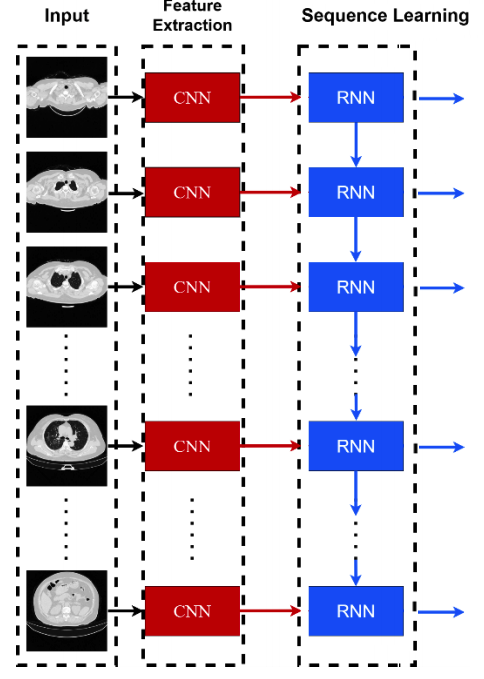}
\caption{The CNN-RNN model}
\label{cnn_rnn}
\end{figure*}

The baseline approach is a CNN-RNN architecture, as shown in Figure \ref{cnn_rnn}. At first all input CT scans are padded to have length $t$ (i.e., consist of $t$ slices). The whole (unsegmented) sequence of 2-D slices of a CT-scan are fed as input to the CNN part. Thus the CNN part performs local, per 2-D slice, analysis, extracting features mainly from the lung regions. The target is to make diagnosis using the whole 3-D CT scan series, similarly to the annotations provided by the medical experts. The RNN part provides this decision, analyzing the CNN features of the whole 3-D CT scan, sequentially moving from slice $0$ to slice $t-1$. The outputs of the RNN part feed the output layer -with 2 units- that uses a softmax activation function and provides the final COVID-19 diagnosis. 

In this way, the CNN-RNN network outputs a probability for each CT scan slice; the CNN-RNN is followed by a voting scheme that makes the final decision; the voting scheme can be either a majority voting or an at-least one voting (i.e., if at least one slice in the scan is predicted as COVID-19, then the whole CT scan is diagnosed as COVID-19, and if all slices in the scan are predicted as non-COVID-19, then the whole CT scan is diagnosed as non-COVID-19).

%There are two modes of producing the input data to our model. In the first, segmentation of each 2-D slice is performed so as to detect the lung regions and the resulting segmented image constitutes the input to the CNN. 

%\subsection{Latent Variable Analysis and Visualization}

%In the proposed methodology we extract and further analyze, through clustering, the, say $L$, neuron outputs of the dense layer of the trained COV19-RACNet.  These latent variables carry high level, semantic information, which is used to generate the final classification at the output layer. We choose to discard the output layer and perform unsupervised analysis of these variables, so as to generate a representation that can provide better visualization of the diagnosis making procedure.   

%Let us assume that we feed the presented architecture with the training dataset. For each CT scan 3-D input $k$, we extract the $L$ neuron outputs of the dense layer, forming a vector ${\textbf{v}}(k)$. In total, we get:

\subsection{Pre-Processing \&  Implementation Details}

At first, CT images were extracted from DICOM files. Then, the voxel intensity values were clipped using a window/level of $350$ Hounsfield units (HU)/$-1150$ HU and normalized to the range of $[0, 1]$. %\textcolor{red}{To reduce the influence of vascular structure and boost the signals of the lesions, a maximum intensity projection (MIP) algorithm was also applied to each of the slices. }
%Data augmentation was also performed, including random rotation in [-10\degree, 10\degree] and horizontal flip. 
%\cite{li2020artificial,zheng2020deep,huang2020serial} to extract ROIs, i.e., lung areas in each CT scan slice.

Regarding implementation of the proposed methodology: i) we utilized ResNet50 as CNN model,  stacking on top of it a global average pooling layer, a batch normalization layer and dropout (with keep probability 0.8); ii) we used a single one-directional GRU layer consisting of 128 units as RNN model. The model was fed with 3-D CT scans composed of the CT slices; each slice was resized from its original size of $512 \times 512 \times 3$ to  $224 \times 224 \times 3$. As a voting scheme, we used the at-least one.

Batch size was equal to 5 (i.e, at each iteration our model processed 5 CT scans) and the input length 't' was 700 (the maximum number of slices found across all CT scans). Softmax cross entropy was the utilized loss function for training the model. Adam optimizer was used with learning rate   $10^{-4}$. Training was performed on a Tesla V100 32GB GPU.

\section{Experimental Results}

This section describes a set of experiments evaluating the performance  of the baseline approach.

Table \ref{3dcnn_rnn} shows the performance of the network over the validation set, after training with the training dataset, in terms of macro F1 score. The macro F1 score is defined as the unweighted average of the class-wise/label-wise F1-scores, i.e., the unweighted average of the COVID-19 class F1 score and of the non-COVID-19 class F1 score.

The main downside of the model is that there exists only one label for the whole CT scan and there are no labels for each CT scan slice. Thus, the presented model analyzes the whole CT scan, based on information extracted from each slice.

\begin{table}[t]
\caption{Performance of the CNN-RNN network}
\label{3dcnn_rnn}
\centering
\scalebox{1.}{
\begin{tabular}{|c|c|}
\hline
Method   &  \multicolumn{1}{c|}{'macro' F1 Score}\\
 \hline
 \hline
ResNet50-GRU   &   0.70  \\
\hline
\end{tabular}
}
\end{table}

\section{Conclusions and Future Work}

In this paper we have introduced  a new large database of chest 3-D CT  scans, obtained in various contexts and consisting of  
different numbers of CT slices. We have also developed a deep neural network, based on a CNN-RNN architecture and used it for  COVID-19 diagnosis on this database. 

%In the  Introduction we provided a detailed description of  the medical significance and impact of chest CT analysis for COVID-19 diagnosis and analyzed the problems of the existing relevant data sets, which led us to aggregate the new database and make it public for research purposes. A large number of significant related references is included in paper's bibliography. The use of current deep learning methods for analysis of existing data sets was referenced in the Related Work Section, also introducing the novelties of our proposed deep neural architecture for COVID-19 diagnosis.  A description and comparison of the new database with reference to the existing ones was presented then. The pre-processing steps and the detailed analysis of the proposed approach was described next. The achieved performance and novel characteristics of our methodology  were illustrated through  a large experimental and ablation study study.  

The scope of the paper is to present a baseline scheme regarding the performance that can be achieved based on analysis of the COV19-CT-DB database.  

The model presented in the paper will be the basis for future expansion towards more transparent modelling of COVID-19 diagnosis.    

{\small
\bibliographystyle{ieee_fullname}
\bibliography{egbib}
}

\end{document}